\documentclass[a4paper]{spie}  
\usepackage[]{graphicx}
\usepackage{amssymb,amsmath,psfrag,url}

\newcommand{\curl}{\, \mathbf{curl}\, } 
\newcommand{\mdiv}{\, \mathbf{div}\, }
\newcommand{\Dint}{\mathrm{\Omega_{\mathrm{int}}}}
\newcommand{\Dext}{\mathrm{\Omega_{\mathrm{ext}}}}
\newcommand{\znum}{\mathbb{Z}}

\newcommand{\rnum}{\mathbb{R}} 
\newcommand{\Field}[1]{{\boldsymbol{#1}}}

\title{Finite-Element Simulations of Light Propagation through Circular Subwavelength Apertures}

\author{
Sven Burger,\supit{\,ab}
Bernd H. Kleemann,\supit{\,c}
Lin Zschiedrich,\supit{\,ab}
Frank Schmidt\supit{\,ab}
\skiplinehalf
\supit{a}
Zuse Institute Berlin,
Takustra{\ss}e 7,
D\,--\,14\,195 Berlin,
Germany
\smallskip\\
\supit{b}
JCMwave GmbH,
Haarer Stra{\ss}e 14a,
D\,--\,85\,640 Putzbrunn,
Germany
\smallskip\\
\supit{c}
Carl Zeiss AG, Carl-Zeiss-Stra{\ss}e 22,
D\,--\,73\,447 Oberkochen,
Germany
}
\authorinfo{
Corresponding author: S. Burger\\
URL: http://www.zib.de/Numerik/NanoOptics/\\
Email: burger@zib.de
}

\begin{document}
\maketitle
\noindent
This paper will be published in Proc.~SPIE Vol. {\bf 7366}
(2009) 736621,  
({\it Photonic Materials, Devices, and Applications III, Ali Serpenguzel, Editor})
and is made available 
as an electronic preprint with permission of SPIE. 
One print or electronic copy may be made for personal use only. 
Systematic or multiple reproduction, distribution to multiple 
locations via electronic or other means, duplication of any 
material in this paper for a fee or for commercial purposes, 
or modification of the content of the paper are prohibited.

\begin{abstract}
Light transmission through circular subwavelength 
apertures in metallic films with surrounding nanostructures is investigated numerically.
Numerical results are obtained with a frequency-domain finite-element method.
Convergence of the obtained observables to very low levels of numerical error is demonstrated.  
Very good agreement to experimental results from the literature is reached, and 
the utility of the method is demonstrated in the investigation of the influence of 
geometrical parameters on enhanced transmission through the apertures.  
\end{abstract}

\keywords{subwavelength aperture, 3D Maxwell solver, finite-element method, nanophotonics, plasmonics}

\section{Introduction}
Experiments by Lezec {\it et al.} investigating light transmission through circular subwavelength 
apertures in metallic films with surrounding nanostructures have revealed surprising 
effects of enhanced transmission and a collimated beam of transmitted light.\cite{Lezec2002a}
The mechanisms involved in the transmission spectra can  
be understood using 2D models and approximative methods\cite{MartinMoreno2003a,GarciaVidal2003a}. 
However, for a fully quantitative 
understanding and from the point of view of optics design, accurate numerical simulations 
of Maxwell's equations for the 3D, cylindrically symmetric problem are desired.
Popov {\it et al.} have developed a well converging field expansion method with a  
Fourier-Bessel basis in order to simulate similar setups.\cite{Popov2005a,Popov2005b,Popov2005c}
To our knowledge results on a simulation of the experiment of Lezec {\it et al.} with this method 
have not been published. 
Baida {\it et al.} have investigated the setting numerically using a finite-difference 
time-domain method (FDTD), however, an agreement of simulation results with the experiment 
is not reached with this method.\cite{Baida2003a} 

We report on a dedicated finite-element method for the simulation task at hand. 
With this, accurately converged results can be obtained in relatively short computation times. 
We discuss simulations of the experimental setup\cite{Lezec2002a} and present 
simulation results agreeing very well with results of the experiment.

The paper is structured as follows:
Section~\ref{subsection_experiments} reviews the general field of light transmission through 
periodic and isolated subwavelength apertures and motivates the need for accurate simulation 
tools. 
The investigated setup of a single aperture surrounded by circular grooves is 
defined in Section~\ref{subsection_setup}.
Section~\ref{subsection_numerical_method} introduces the finite-element solver JCMsuite used in 
this study. 
Numerical results are presented in Chapter~\ref{section_results} 
including a convergence study for the investigated setup in Section~\ref{subsection_convergence}, 
results on the experimental setup of Lezec {\it et al.} and on the setup investigated by 
Baida {\it et al.} in Section~\ref{subsection_beaming} ,
and results on the influence of a geometrical parameter on  
transmission spectra through the subwavelength apertures in Section~\ref{subsection_geo_dependence}. 
 Chapter~\ref{section_conclusions} concludes the paper.

\section{Background}
\subsection{Enhanced transmission through isolated and periodic subwavelength apertures}
\label{subsection_experiments}

In 1997 experiments were reported \cite{Ebbesen1998a} claiming surprisingly high transmission of light 
through 2D-periodic sub-wavelength hole arrays drilled in a metallic sheet of silver. 
The enhancement was reported to be orders of magnitude larger as compared with the transmission 
of one single hole of same size predicted by standard aperture theory.\cite{Bethe1944a} 
The explanation of the experiments was mainly based on the existence 
of surface plasmons polaritons (SPPs) that are excited by the incident light field.
The experiment has created a lot of attention, and the perception of surface plasmons has 
been widely supported by further investigation. 
However, in 2004 Lezec and Thio 
report on the experimental observation of both enhancement 
and suppression in the transmission spectra of hole arrays in metallic films.\cite{Lezec2004a} 
Their work suggests that a lateral interference phenomenon of an evanescent wave with the incident light
is the driving force in shaping 
the transmission spectra. 
Especially, a suppression in the transmission spectrum is not
be explained by the SPP model  which 
predicts only enhancement with respect to the single-hole case. 
Enhanced transmission with a similar signature is also reported for hole arrays in tungsten 
which does not support SPPs in the investigated regime. 
The authors further report on a transmission enhancement factor 
not larger than 7, 
in contrast to the often-quoted enhancement factor of order~1000.\cite{Lezec2004a,note_enhancement_factor}
The explanation of the transmission spectra with interference phenomena has been supported also 
by a study showing that the transmission peak positions are determined by 
the period of the hole array and are material independent.\cite{Garcia2006a} 
Flammer {\it et al.} studied transmission through a single slit surrounded 
by linear grating structures.\cite{Flammer2007a} 
Also this work confirms the explanation through an interference phenomenon. 
However, in this case 
it has been found that the surface wave is an SPP, and not an evanescent wave. 
The theoretical model of this work shows that a SPP is launched along the metal surface, 
while interference of the SPP with the incident light along with resonant cavity effects 
give rise to suppression and enhancement in transmission. 
In another work it has even been shown  that there are no holes necessary for enhanced transmission
through thin metallic films provided the film is modulated.\cite{Bonod2003a}

This story of different interpretations of experimental results points out that 
approximate models often trigger in-depth physical understanding of the ongoing physics. 
However, when different models lead to partly contradictive conclusions, a comparison with 
rigorous simulations has to be performed. 
These methods need to consider the full
wavelength dependent material dispersion behaviour, must be able to reach a sufficient 
spatial resolution of the structures and should have a good convergence rate 
to get fast and accurate results. 

\subsection{Investigated Setup}
\label{subsection_setup}

\begin{figure}[htb]
\centering
\psfrag{h1}{\sffamily $h$}
\psfrag{h2}{\sffamily $t$}
\psfrag{p}{\sffamily $p$}
\psfrag{w1}{\sffamily $w$}
\psfrag{w2}{\sffamily $d$}
\psfrag{x}{\sffamily $r$}
\psfrag{y}{\sffamily $y$}
\fbox{\hspace{0.5cm} \includegraphics[width=0.7\textwidth]{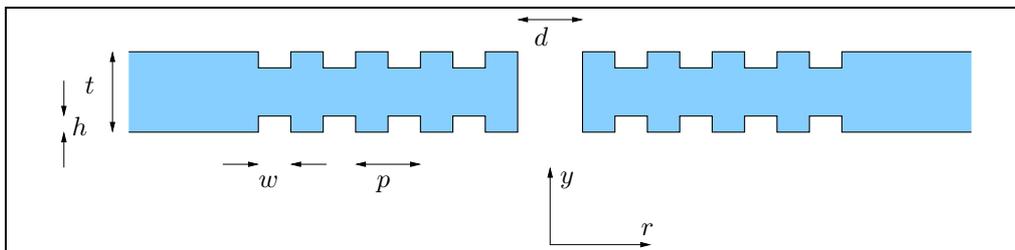}\hspace{0.5cm}}
\caption{Schematics of the experimental setup: silver sheet (light blue) of thickness $t$ is perforated by a 
cylindrical hole of diameter $d$, surrounded by four circular grooves of width $w$, depth $h$
and distance $p$. The setup is rotationally symmetric around the $y$-axis. 
An illuminating plane wave is incident in $y$-direction. 
}
\label{schema_lezec_xy}
\end{figure}

The investigated setup is relatively simple: 
A plane wave under normal incidence illuminates a silver sheet which is perforated by a small circular 
hole, surrounded by assist features. The geometry is symmetric under rotation around the axis of the 
hole. 
Figure~\ref{schema_lezec_xy} schematically shows the geometrical setup,\cite{Lezec2002a} 
the parameters are given in Table~\ref{table_layout_definitions}.
For the relative electric permittivity of silver, $\varepsilon_\mathrm{r}$, we assume a Drude model,  
$\varepsilon_\mathrm{r} = 1 - \omega_\mathrm{p}^2/(\omega^2+i \, \omega \,\omega_\mathrm{c})$, 
with plasma frequency $\omega_\mathrm{p}$ and collision frequency
$\omega_\mathrm{c}$ as given in Table~\ref{table_layout_definitions}, and frequency $\omega = c_0 2\pi/\lambda_0$. 
The geometry and material parameters correspond to the parameters of Ref.\cite{Baida2003a} (structure {\it CAG}),
investigating the same experimental setup.\cite{Lezec2002a} 

\begin{figure}[htb]
\centering
\fbox{
\includegraphics[width=0.98\textwidth]{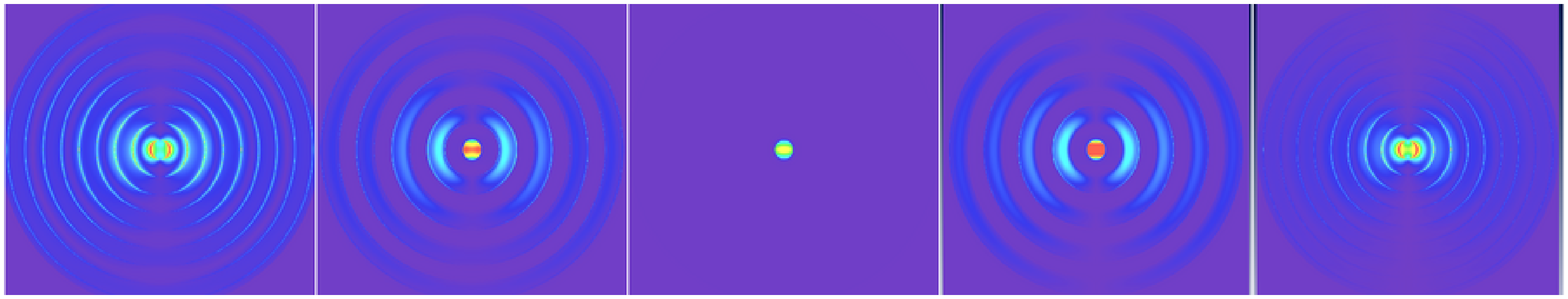}
}
\caption{
Pseudocolor images of the magnitude of a simulated electric field, $\beta |\Field{E}|$, 
in five cross-sections parallel to the surface of the silver film, at different heights $y$. 
From left to right: $y=-151\,$nm (1\,nm below the surface on the entrance side), 
$y=-91\,$nm (1\,nm below the bottom of the grooves on the entrance side), 
$y=0\,$nm (at the center of the film), 
$y=91\,$nm (1\,nm above the bottom of the grooves on the exit side), 
$y=151\,$nm (1\,nm above the surface on the exit side).
The pseudocolor images (red: high magnitude, blue: low magnitude) are scaled by a factor $\beta$
relative to the amplitude of the incoming plane wave
(from left to right: $\beta = 4\%, 8\%, 8\%, 20\%, 10\%$).
The parameters of the simulation are given in Table~\ref{table_layout_definitions} (set 5).
[{\it See original paper for better image resolution.}]
}
\label{field_plots}
\end{figure}

In sets of successive simulations either numerical parameters are scanned in order to investigate the 
convergence of the obtained discrete solutions, or physical parameters like the wavelength or geometry parameters 
are scanned in order to obtain, e.g., the spectral response of the structure.
The angular spectrum of transmitted light is investigated by evaluating Fourier transforms 
of the computed field at the boundaries of the computational domain. 
For an analysis of the field distributions, these can also be exported and visualized. 
Figure~\ref{field_plots} shows an example where a simulated near field is visualized in different 2D sections 
through the 3D field distribution. 
The sections clearly resemble the cylindrical  geometry of the structure.

\begin{table}[h]
\begin{center}
\begin{tabular}{|r|r|r|r|r|r|}
\hline
&{\bf  set 1}&{\bf set 2}&{\bf set 3}&{\bf set 4}&{\bf set 5}\\ 
\hline
\hline
\multicolumn{6}{|l|}{Geometry parameters} \\
\hline
$d$ & \multicolumn{2}{|l|}{330\,nm}&  \multicolumn{3}{|l|}{300\,nm}\\
\hline
$t$ & \multicolumn{5}{|l|}{300\,nm} \\
$h$ & \multicolumn{5}{|l|}{60\,nm} \\
\hline
$p$ & \multicolumn{3}{|l|}{600\,nm}&550\,nm - 650\,nm &  565\,nm\\
\hline
$w$ &  \multicolumn{5}{|l|}{$p/2$} \\
\hline
\hline
\multicolumn{6}{|l|}{Incident spectrum} \\
\hline
$\lambda$ & 670\,nm &400\,nm -- 1100\,nm & \multicolumn{2}{|l|}{500\,nm -- 800\,nm} & 631.4\,nm \\
\hline
\hline
\multicolumn{6}{|l|}{Material properties} \\
\hline
$\omega_\mathrm{p}$ & \multicolumn{5}{|l|}{1.374e16 s$^{-1}$}\\
$\omega_\mathrm{c}$ & \multicolumn{5}{|l|}{3.21e13 s$^{-1}$}\\
\hline
\end{tabular}
\caption{
Five different sets of used parameter settings, for geometry parameters compare Fig.~\ref{schema_lezec_xy}.
Please note that we investigate settings with two different values for the central hole diameter $d$
because we compare to results from a numerical study\cite{Baida2003a} using $d=330\,$nm and we compare 
to experimental results\cite{Lezec2002a}  where the setting $d=300\,$nm was reported. 
}
\label{table_layout_definitions}
\end{center}
\end{table}

\subsection{Numerical Method}
\label{subsection_numerical_method}

For solving the time-harmonic Maxwell's equations describing light scattering we 
use a dedicated finite-element method for nanooptics simulations:
The finite-element package  {\it JCMsuite} is a joint development of 
the Zuse Institute Berlin and JCMwave. 
This solver has been successfully applied to a wide 
range of electromagnetic field computations including
metamaterials,\cite{Enkrich2005a}
photonic crystal fibers,\cite{Couny2007a} 
nearfield-microscopy,\cite{Kalkbrenner2005a}
and optical microlithography.\cite{Burger2007om}
The solver is also used for pattern reconstruction in 
optical metrology.\cite{ScLaDe07}
The performance of our methods has been pointed out in several 
benchmarks.\cite{Burger2005bacus,Burger2008bacus}
The main ingredients for an accurate performance are adaptive methods, based on 
goal-oriented error estimators, higher-order vector elements, and fast direct and 
iterative numerical methods for solving matrix equations.\cite{Pomplun2007pssb} 

In this contribution we apply the solver for light scattering computations on 
rotationally symmetric geometries, included in the programme package {\it JCMsuite}.
Assuming harmonic time-dependence of the electric field with frequency $\omega$, 
$\Field{E}_{\mathrm{t}} = \Field{E}\exp(-i\omega t)$,
Maxwell's equations can be written as
\begin{subequations}
\label{Eqn:THMax}
\begin{eqnarray}
\curl \mu^{-1} \curl \Field{E} -\omega^{2}\varepsilon \Field{E} &  = & 0 \quad ,\\
\mdiv \epsilon \Field{E} & =& 0 \quad .
\end{eqnarray}
\end{subequations}
The spatial distribution of the tensorial permittivity $\varepsilon$ and permeability $\mu$ 
can be chosen (nearly) arbitrarily in the computational window $\Dint \subset \rnum^{3}.$ 
The scattering problem is defined as follows: 
Given an incoming electric field $\Field{E}_{\mathrm{inc}}$ compute the total
electric field $\Field{E}$ (satisfying~\eqref{Eqn:THMax} in $\rnum^{3}$)
such that the scattered field 
$\Field{E}_{\mathrm{sc}} = \Field{E}-\Field{E}_{\mathrm{inc}}$ 
defined on the exterior domain $\Dext = \rnum^{3} \setminus \Dint$ is purely outward radiating.
 
Using the transformation rules for differential forms with Jacobian matrix $\mathrm{J}$ of 
the transformation $(r, y, \phi) \mapsto (x, y, z)$ and
$\curl_{*} = (\partial_{r}, \partial_{y}, \partial_{\varphi})^{\mathrm{t}} \times$, 
$\Field{E}_{*} = \mathrm{J}^{\mathrm{t}} \Field{E}$, 
$\varepsilon_{*} = |\mathrm{J}|\mathrm{J}^{\mathrm{-1}} \varepsilon \mathrm{J}^{\mathrm{-t}}$ and  
$\mu_{*} = |\mathrm{J}|\mathrm{J}^{\mathrm{-1}} \mu \mathrm{J}^{\mathrm{-t}}$, Maxwell's equations 
in cylinder coordinates $(r, y, \varphi)$ read as
\begin{subequations}
\begin{eqnarray}
\curl_{*} \mu^{-1}_{*} \curl_{*} \Field{E}_{*} -\omega^{2} \varepsilon_{*} \Field{E}_{*} & = &  0 \quad ,\\
\mdiv_{*} \epsilon_{*} \Field{E}_{*} &=& 0 \quad .
\end{eqnarray}
\end{subequations}
We assume now that the material distribution is cylindrically symmetric, i.e., 
$\varepsilon_{*} = \varepsilon_{*}(r, y)$,
$\mu_{*} = \mu_{*}(r, y)$,
and we expand
the transformed incoming field 
$\Field{E}_{\mathrm{inc}, *} = \mathrm{J}^{\mathrm{t}} \Field{E}_{\mathrm{inc}}$
in a Fourier series,
\[
\Field{E}_{\mathrm{inc}, *}(r, y, \varphi) = \sum_{n=-\infty}^{n=\infty} \Field{e}_{\mathrm{inc}, n} (r, y)e^{in\varphi} \quad .
\] 
With this, the 3D Maxwell's equations separate into 2D problems for each $n \in \znum,$
\begin{equation}
\label{Eqn:SepCyMax}
\curl_{*, n} \mu^{-1}_{*} \curl_{*, n} \Field{e}_{n} -\omega^{2} \varepsilon_{*} \Field{e}_{n} =  0 \quad ,
\end{equation}
with $\curl_{*, n}=(\partial_{r}, \partial_{y}, i n)^{\mathrm{t}} \times.$ 
Equation~(\ref{Eqn:SepCyMax}) in its weak formulation 
is solved several times for increasing $|n|$. 
The computation is stopped automatically when the incoming modes  
$ \Field{e}_{\mathrm{inc}, n}$ become sufficiently small.

\section{Numerical Results}
\label{section_results}
\subsection{Convergence}
\label{subsection_convergence}
We demonstrate sound convergence properties of our finite-element solver 
by performing several computations of the 
same physical setup with increasing numerical resolution. 
The physical parameters for this convergence study are given in Table~\ref{table_layout_definitions} (parameter set 1), 
the wavelength of the incident 
plane wave corresponds to a transmission peak for this parameter setting.

\begin{figure}[htb]
\centering
\fbox{
a)
\includegraphics[width=0.4\textwidth]{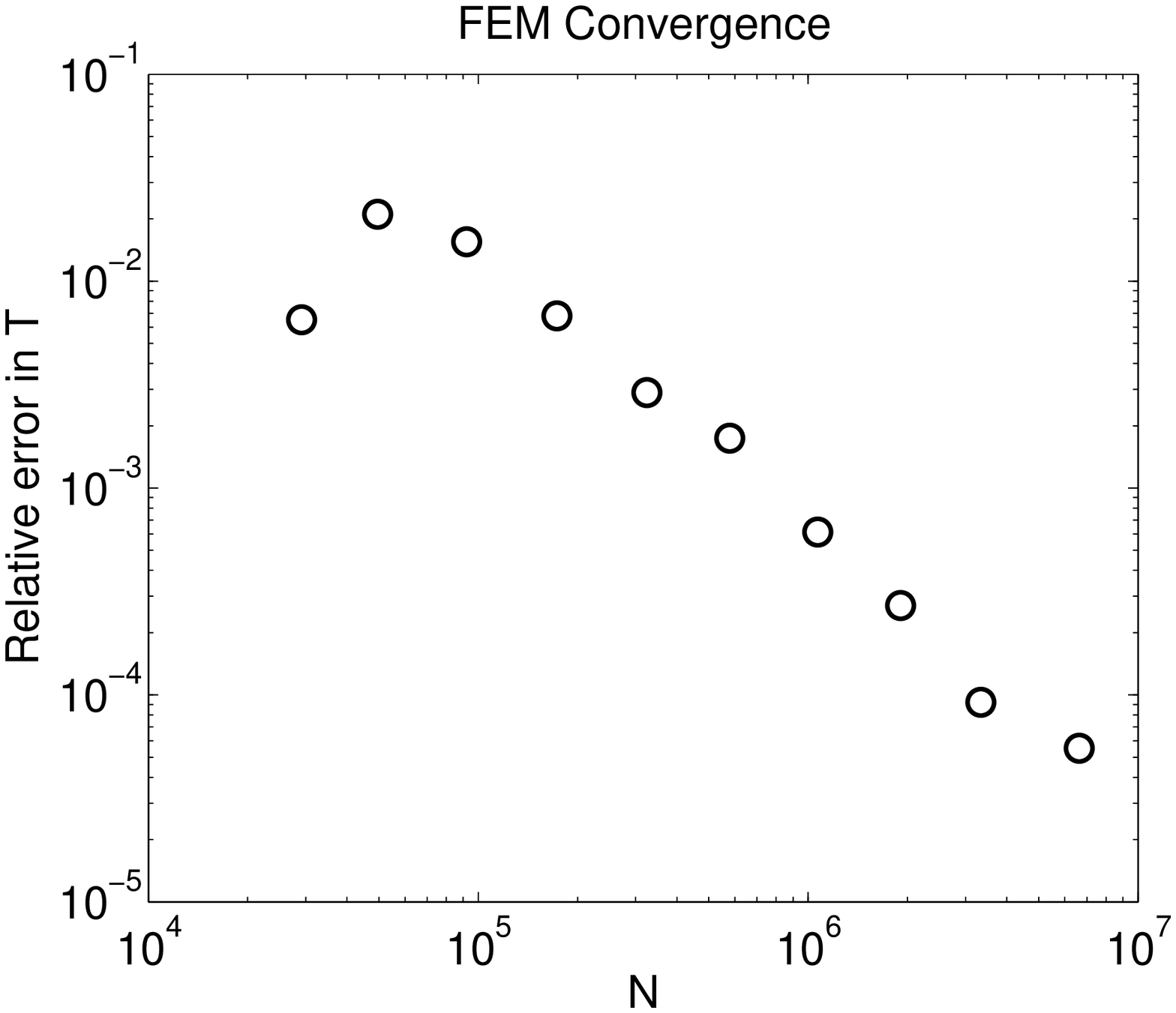}
\hspace{1cm}
b)
\includegraphics[width=0.4\textwidth]{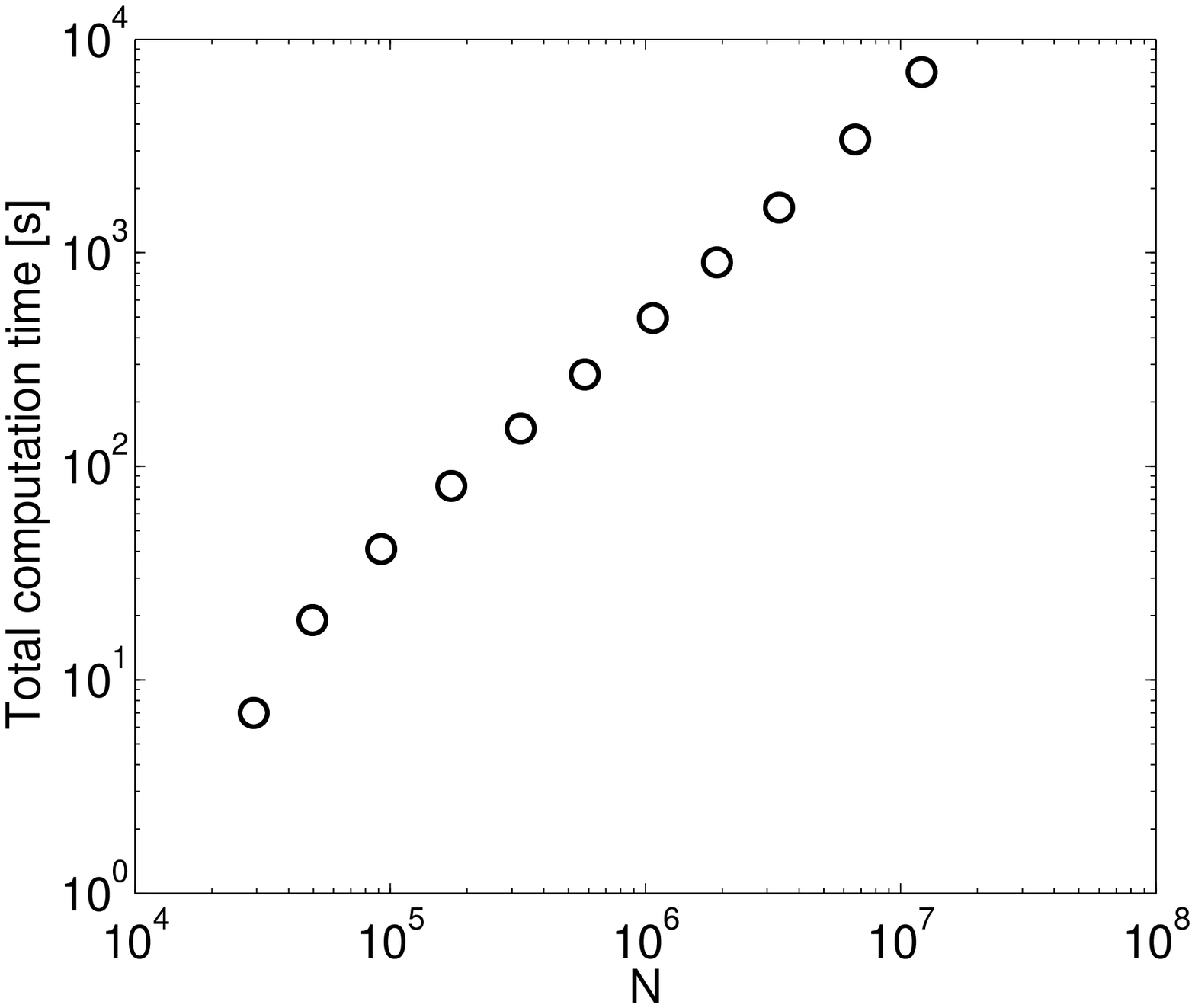}
\hspace{0.2cm}}
\caption{
a) Convergence plot: Relative error of computed transmission through a nanoaperture in dependence 
on number of unknowns $N$ of the FEM problem. 
b) Computational effort: Computation time in seconds (single-thread computation on a standard workstation) in dependence 
on $N$ (corresponding to the computations displayed in a). 
}
\label{convergence}
\end{figure}

The physical observable of interest in this case is the far field intensity transmitted through the aperture, $T$, 
detected at a specific detection angle. $T$ is obtained from the FEM near field solution. 
The far field intensity $T(N)$ computed from a numerical approximation with $N$ degrees 
of freedom, $\Field{E}_N$, in general differs from  $T$ computed from the exact solution. 
The relative error of $T(N)$ is defined as $\Delta T(N) = |T(N)-T_\mathrm{qe}|/|T_\mathrm{qe}|$, where 
$T_\mathrm{qe}$ is the intensity computed from the quasi-exact solution, i.e., 
from a solution on a finer mesh than the meshes of the solutions corresponding to $T(N)$. 
Indeed, it would be desirable to compare $T(N)$ to an analytical solution. 
However, for problems where an analytical solution is not available, the quasi-exact solution 
is used as a makeshift. 
To validate the results we have further tested that results of our method
converge towards results obtained from Mie-Theory (for scattering off a sphere), and we 
have further quantitatively reproduced published results on light transmission through circular holes obtained from 
a Fourier-Bessel modal method.\cite{Popov2005a}

Figure~\ref{convergence} shows the dependence of the relative error $\Delta T(N)$ on the 
number of degrees of freedom in the numerical approximation, $N$.
The detection angle is in normal direction to the silver film in this case. 
Increasing the finite-element polynomial degree and refining the mesh, 
both leads to an increase in $N$.
And an increase in $N$ leads -- when the convergent regime is reached -- to 
a decrease of the relative error of the numerical approximation. 
For the displayed results we have used finite-elements of second polynomial order,   
and we have used up to ten successive adaptive mesh refinement steps. 
Figure~\ref{convergence} further displays the computational effort in terms of computation time on a  
standard workstation. 
The software has multi-processing capabilities, however, for a better 
comparability we have used a single core of a cpu only for these computations. 
The convergence plot shows that for achieving a relative error of, e.g., one percent a moderate 
number of unknowns of about $ N = 10^5$ is needed, corresponding to computation times well below one minute on a 
standard personal computer (PC). 

\subsection{Enhanced transmission and beaming effect}
\label{subsection_beaming}

\begin{figure}[htb]
\centering
\fbox{
a)\includegraphics[width=0.46\textwidth]{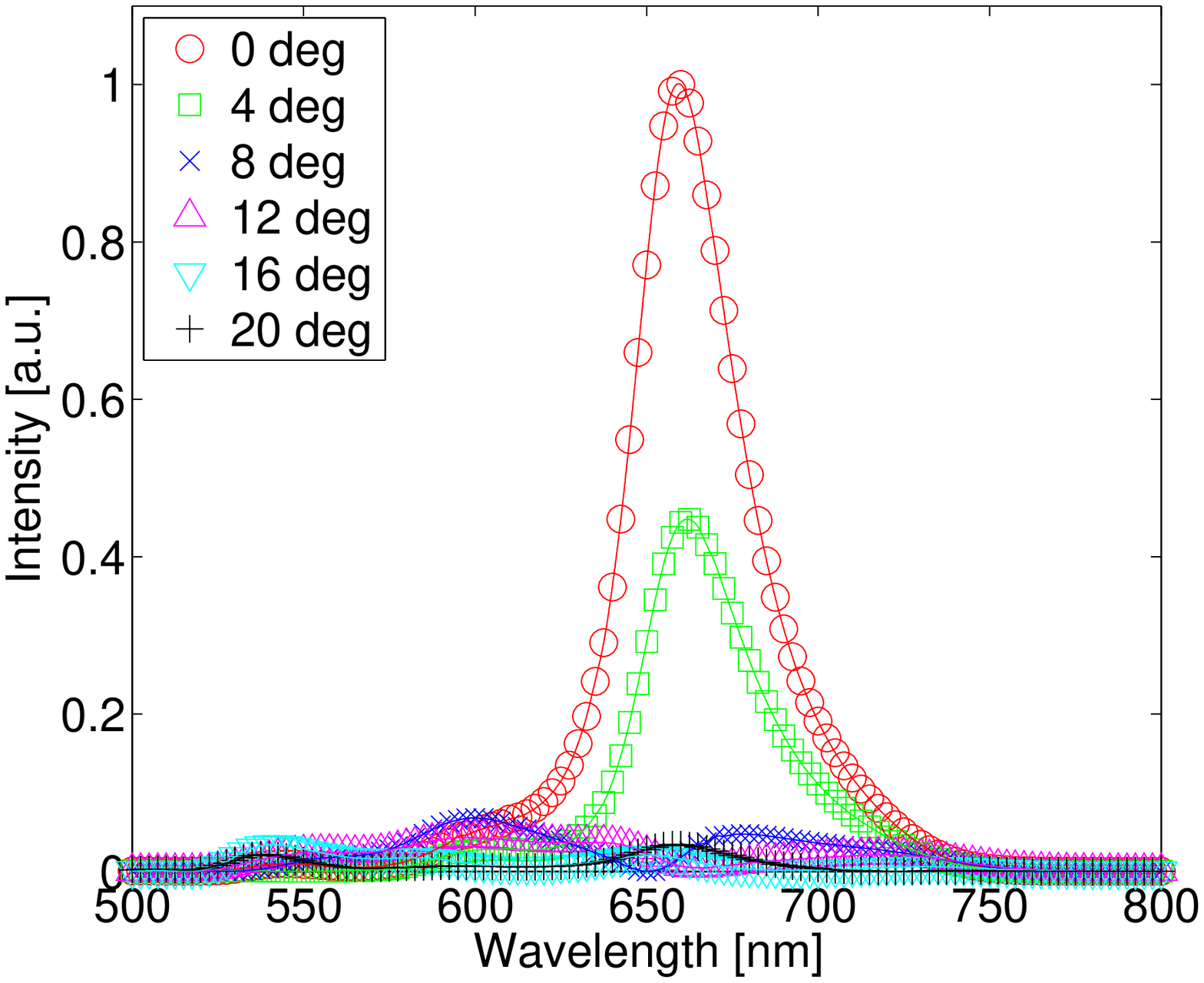}
b)\includegraphics[width=0.46\textwidth]{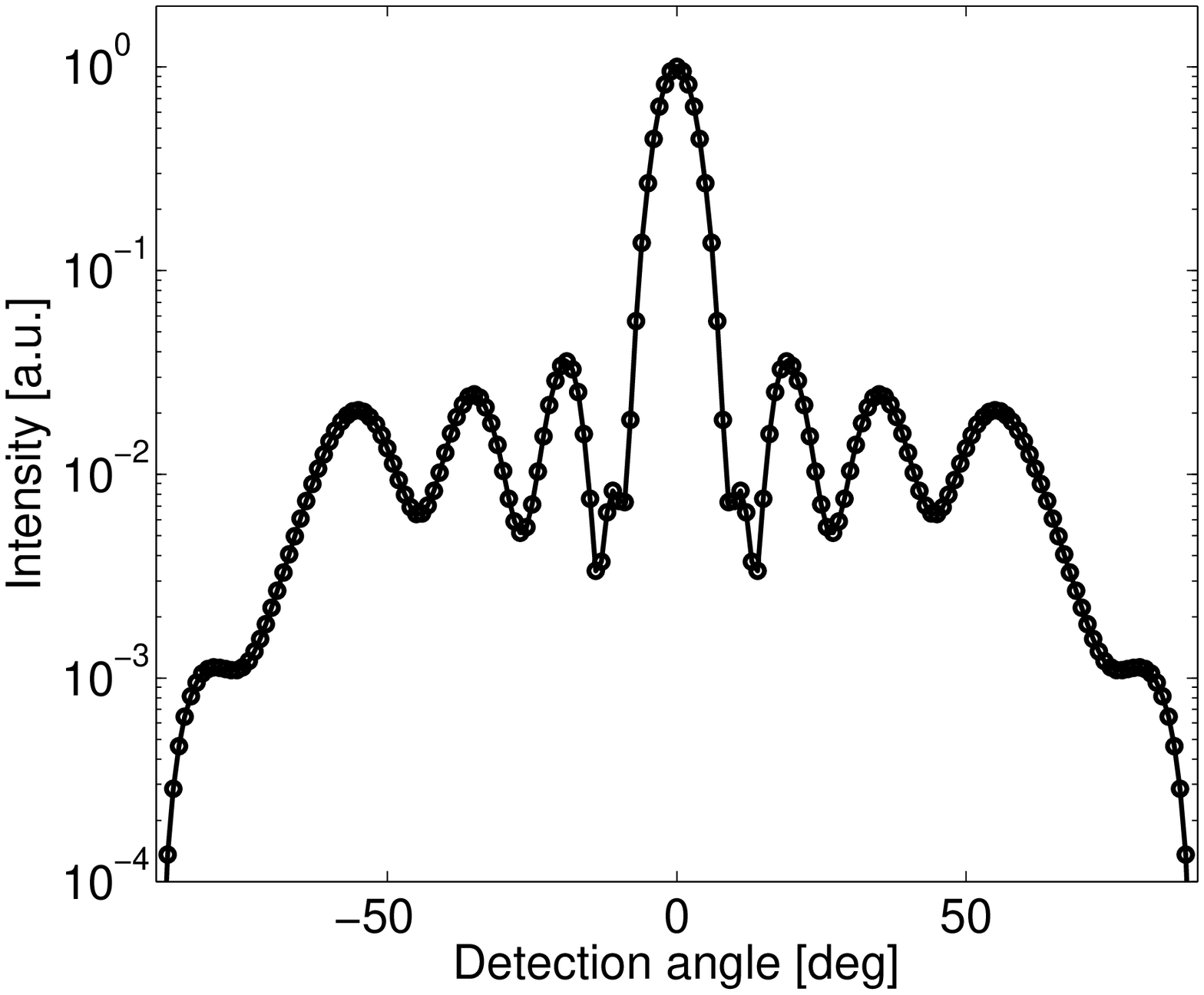}
}
\caption{
a) Intensity spectrum of transmitted light for different detection angles. 
b) Angular intensity spectrum of transmitted light at maximum transmission 
wavelength ($\lambda = 660\,$nm).
}
\label{spectra_lezec}
\end{figure}

In order to numerically investigate the experimental setting of Lezec~{\it et al.} we have performed a wavelength 
scan and recorded far-field transmission into different spatial directions. 
The physical parameters in this case were as defined in  Table~\ref{table_layout_definitions} (parameter set 3).
These parameters correspond to the published data by Lezec~{\it et al.}\cite{Lezec2002a}
Figure~\ref{spectra_lezec} shows the simulated spectra. 
Clearly, enhanced transmission around a peak wavelength of $\lambda=660\,$nm is found for low observation 
angles. 
This is in very good agreement with the experimental findings.\cite{Lezec2002a}

For reproducing the beaming effect observed in the experiment we further compute the far-field pattern 
of transmitted light at wavelength of maximum transmission through the nanoaperture. 
Figure~\ref{spectra_lezec}\,b) shows the corresponding angular spectrum of diffracted light: 
Transmitted light emerges from 
the aperture as a beam with small angular divergence of approximately $\pm 3.5$\,degree (FWHM).
Again these results agree very well with the experimental findings.

\begin{figure}[htb]
\centering
\fbox{\includegraphics[width=0.45\textwidth]{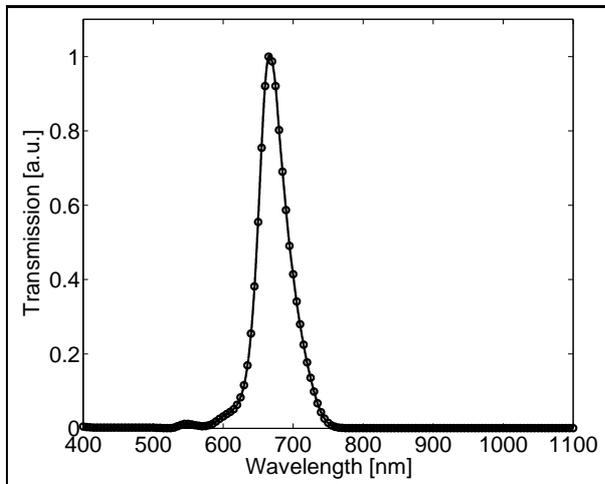}}
\caption{
Intensity spectrum of transmitted light at a detection angle of 0~degree. 
Geometry as defined in  parameter set 1
of Table~\ref{table_layout_definitions}.
Transmission peak around a wavelength of $\lambda = 670\,$nm. 
}
\label{spectrum_baida}
\end{figure}

For comparison to results obtained with a FDTD method 
we also include an intensity spectrum for the geometrical values defined in  
Table~\ref{table_layout_definitions} (parameter set 2).
The geometry and material parameters correspond to the parameters of Ref.\cite{Baida2003a} 
Figure~\ref{spectrum_baida} shows the computed spectrum.
We note that this spectrum is significantly different from the spectrum computed by 
Baida {\it et al.} using FDTD (Fig.~2b in Ref.\cite{Baida2003a}).
The spectrum of Ref.\cite{Baida2003a} is also in discrepancy with the experimental results of 
Lezec  {\it et al.}
In Ref.\cite{Baida2003a} the discrepancy between simulation results and experimental findings is attributed 
to either the difference between the dispersive properties of the real metal used in the experiment and 
the data used in the simulations, or to a possible difference between geometrical parameters used in the 
modeling and those of the real experiment. 
However, our results with the same geometry and material parameters, 
as shown in Figure~\ref{spectrum_baida}, agree well with the experimental findings. 
A slight shift ($\approx 10\,$nm) in transmission peak wavelength is explained by a difference in 
aperture diameter between numerical parameters ($d=330\,$nm) and experimental setting ($d=300\,$nm).  
This suggests that the discrepancy in Ref.\cite{Baida2003a} is rather caused by insufficient convergence 
of the used numerical methods. 

\subsection{Dependence of enhanced transmission on geometrical parameters}
\label{subsection_geo_dependence}

\begin{figure}[htb]
\centering
\fbox{
a)\includegraphics[width=0.46\textwidth]{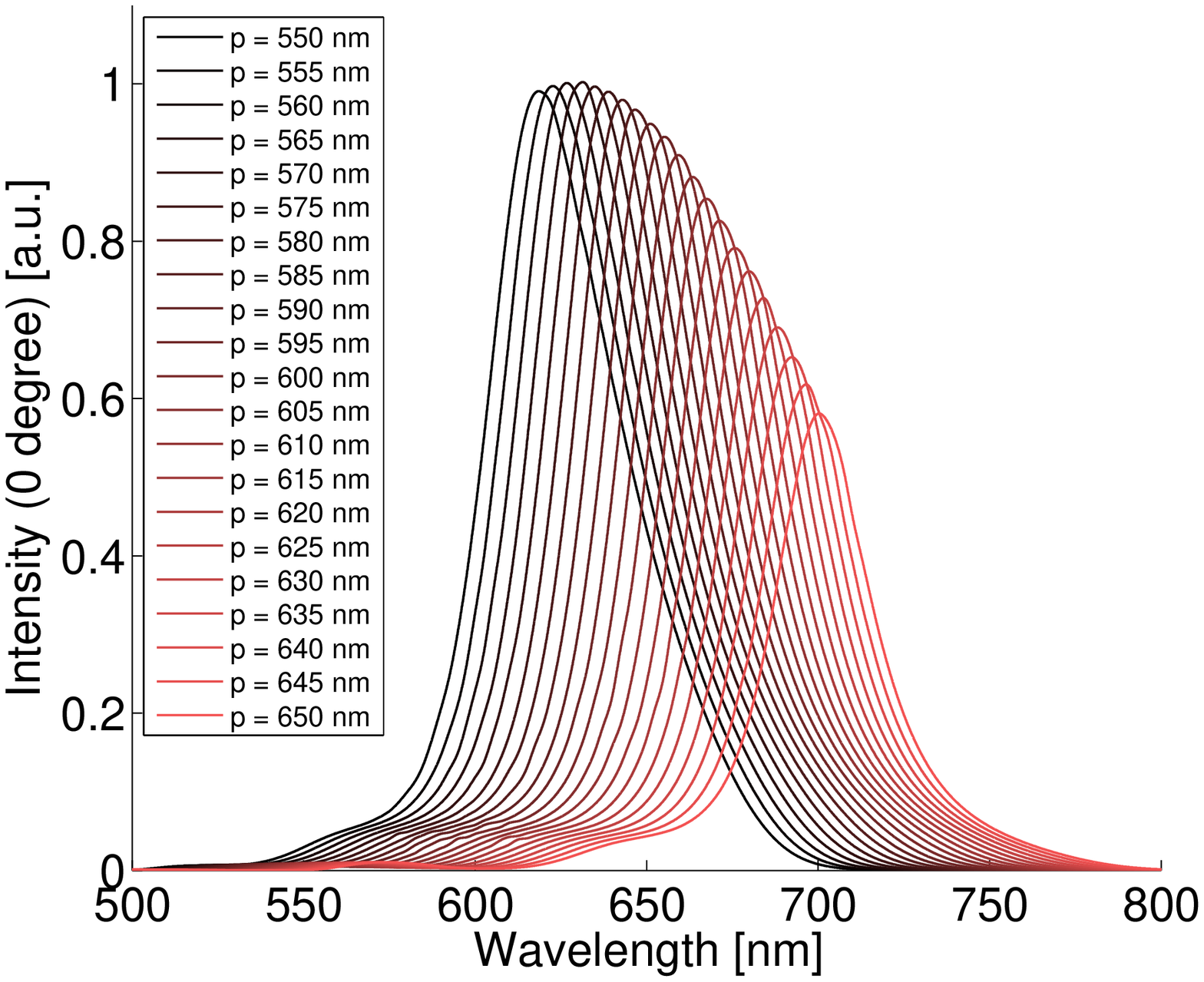}
b)\includegraphics[width=0.46\textwidth]{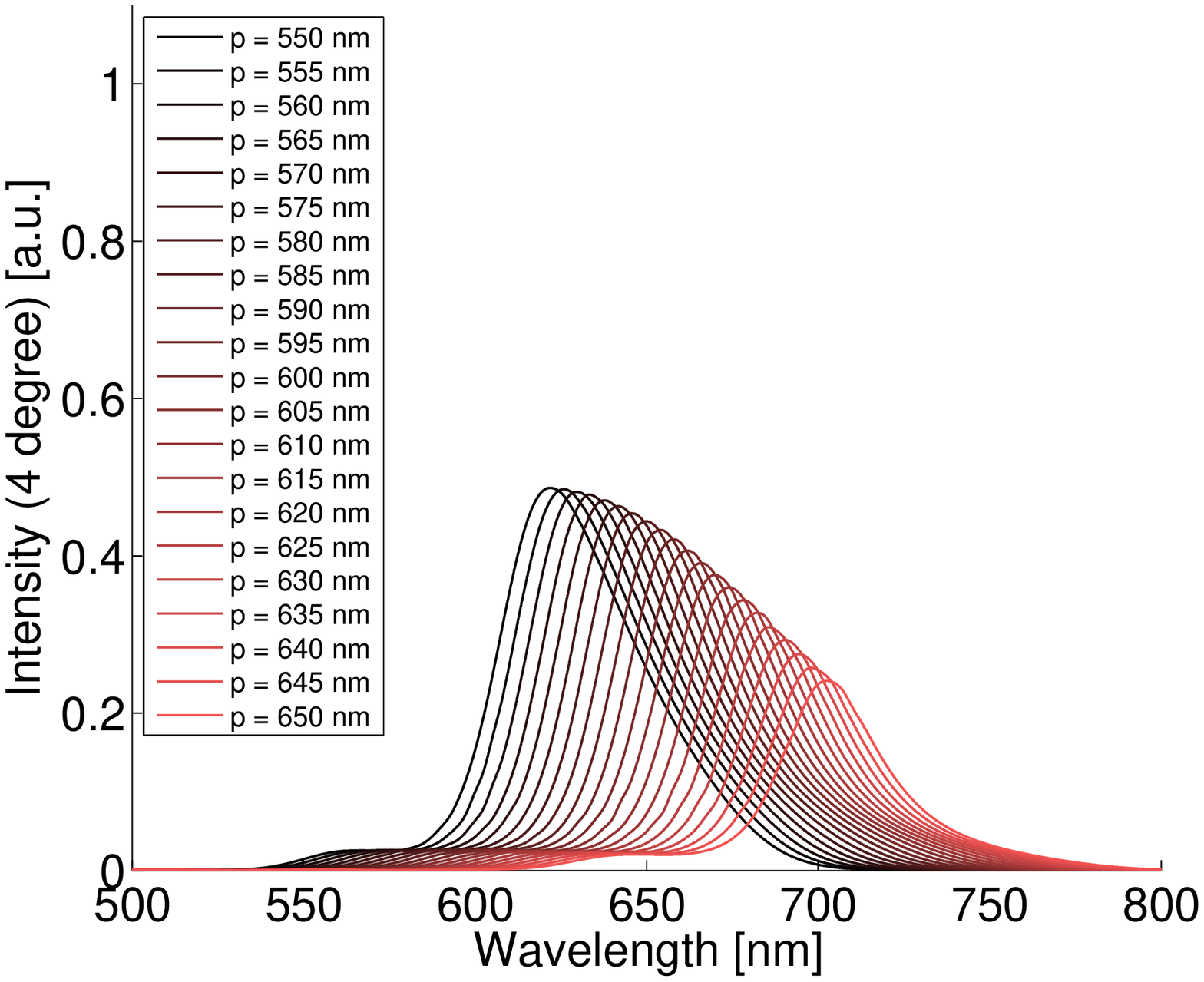}
}
\caption{
Dependence of the transmission spectrum on the spacing $p$ of the 
circular grooves surrounding the nanoaperture.  
Intensity spectra of transmitted light at a detection angle of 0~degree (a), 
resp.~4~degree (b).
The groove spacing is varied between 550\,nm (black, peak at $\lambda \approx 618.5\,$nm)
and 650\,nm (red, peak at $\lambda \approx 700.5\,$nm).
}
\label{spectra_pitches}
\end{figure}

We have 
performed a set of numerical experiments where we simulate transmission spectra for 
different spacings $p$ of the grating structure surrounding the aperture ({\it cf.} Fig.~\ref{schema_lezec_xy}). 
All other parameters of the setup remain fixed to set 4 of Table~\ref{table_layout_definitions}.
As numerical parameters, second order finite-elements and two successive adaptive refinement steps of the 
coarse grids were chosen. 
According to Section~\ref{subsection_convergence} this setting corresponds to 
a relative error of the transmission values of about one percent. 
Figure~\ref{spectra_pitches} shows how the transmission maxima for detection angles 0~degree and 4~degree 
are shifted to the red side of the spectrum with increasing spacing $p$. 
In the investigated regime, the spectral position of the maxima depends nearly linearly on spacing $p$.
The significant shift of the transmission peak demonstrates the strong influence of the 
structures surrounding the nanoaperture, emphasizing the importance of interference effects for 
enhanced transmission through subwavelength apertures. 

A maximum of enhanced transmission is observed at a spacing of $p=565\,$nm and a wavelength 
of $\lambda = 631.4\,$nm.
Figure~\ref{field_plots} shows the near field amplitude in different cross sections through the 3D 
field distribution for these parameters (parameter set 5 of Table~\ref{table_layout_definitions}). 
It is interesting to see how well the field distribution at the exit side of the structure 
resembles the field distribution on the input side, with a relative amplitude scaling of  
(only) about $2.5$.

\section{Conclusions}
\label{section_conclusions}

We have presented a finite-element method for 
accurate and fast, rigorous simulations of light propagation 
in cylindrically symmetric, metallic nano-structures. 
Simulation results of light transmission through structured 
nano-apertures show an excellent agreement with experimental results 
from the literature. 
The method has been used to investigate the strong dependency of 
transmission spectra on a geometrical parameter. 

With the presented finite-element solver, a method is at hand which allows for more detailed studies of light 
transmission through subwavelength apertures, 
and for an efficient and reliable design of metallic nanostructures for specific applications,
e.g., in sensing or for near field microscopy. 

\bibliography{spie}
\bibliographystyle{spiebib}
\end{document}